\documentclass[12pt]{revtex4}
\usepackage{amssymb,epsf}
\begin{document}

\title{Hairy rotating black string in the Einstein-Maxwell-Higgs system}
\author{M. H. Dehghani$^{1,2}$}\email{mhd@shirazu.ac.ir} \author{A. Khodam-Mohammadi$^{1}$}
\address{${1}$. Physics Department and Biruni Observatory, College of Sciences, Shiraz
University, Shiraz 71454, Iran\\
        ${2}$. Institute for Studies in Theoretical
        Physics and Mathematics (IPM), P.O. Box 19395-5531, Tehran, Iran}

\begin{abstract}
We show numerically that the Abelian Higgs field equations in the
background of a four-dimensional rotating charged black string
have vortex solutions. These solutions which have axial symmetry
show that the rotating black string can support the Abelian Higgs
field as hair. We find that one encounters with an electric field
coupled to the Higgs scalar field for the case of rotating black
string. This electric field is due to an electric charge per unit
length, which increases as the rotation parameter becomes larger.
We also find that the vortex thickness decreases as the rotation
parameter grows up. Finally we consider the self-gravity of the
Abelian Higgs field and show that the effect of the vortex is to
induce a deficit angle in the metric under consideration which
decreases as the rotation parameter increases.
\end{abstract}

\maketitle
\section{Introduction}

The classical no hair conjecture, first proposed by Ruffini and Wheeler
states that the only allowed characteristics of stationary black holes are
those associated with the Gauss law, such as mass, angular momentum and
electromagnetic charge \cite{Ruf}. Thus, for example, it follows that black
holes cannot support external scalar fields, since there is no Gauss law for
scalars. This fact is verified in \cite{Sud} for a scalar field minimally
coupled to gravity and in \cite{Saa} for non-minimally coupled scalar field
in asymptotic flat spacetimes. However, recently, it is shown that there
exist scalar hairy black hole if one uses a scalar potential with regions
violating the weak energy condition \cite{Nuc}. In recent year, we are faced
with new black hole solutions in special gauge theories that Einstein's
equation is coupled with some gauge fields, and therefore this conjecture
needs more investigation. For example the Einstein-Yang-Mills black hole
possess a short ranged external non Abelian gauge field and is not uniquely
specified by their mass, angular momentum and electromagnetic charge \cite
{VB}. Although this solution is unstable \cite{un}, it opened up new
possibility that the black holes may have various matter hairs. The global
monopole field is an example of stable hair for asymptotically flat black
hole \cite{Wat}. Also explicit calculations have been carried out which
verify the existence of a long range Nielson-Olesen vortex solution as a
stable hair for a static black hole solution of Einstein's equation in four
dimensions \cite{Gre}. For the extreme black hole of Einstein-Maxwell
gravity, it has been shown that flux expulsion occurs for thick strings
(thick with respect to the radius of horizon), while flux penetration occurs
for thin strings \cite{Cha1,Cha2,Bon1,Bon2}. Of course one may note that
these situations fall out side of the scope of the classical no hair theorem
due to the non trivial topology of the string configuration.

It has recently been shown that these ideas can be extended to the
case of anti de sitter (AdS) and de Sitter (dS) spacetimes. For
asymptotically AdS spacetimes, it has been shown that conformally
coupled scalar field can be painted as hair \cite{Eli2}. Another
asymptotically AdS hairy black hole solution has been investigated
in Ref. \cite{Torii2}. Also it was shown that there exist a
solution to the $SU(2)$ Einstein-Yang-Mills equations which
describes a stable Yang-Mills hairy black hole, that is
asymptotically AdS \cite{Eli1}. Although the idea of
Nielson-Olesen vortices has been first introduced in flat
spacetimes \cite{NO} , but recently it has been extended to (A)dS
spacetimes \cite{Deh1,Ghez1}. The existence of long range
Nielson-Olesen vortex as hair for asymptotically AdS black holes
has been investigated in Refs. \cite{Deh2,Dehjal} for
Schwarzschild-AdS black hole and charged black string.

The presence of a rotation parameter introduces significant
additional difficulties for explicit calculations intended to
investigate the existence of a long range Nielson-Olesen vortex
solution as a hair for a stationary black hole solution
\cite{Ghez2}. In this paper we study the Abelian Higgs hair in a
four dimensional rotating charged black string that is a
stationary model for Einstein-Maxwell equations in cylindrical
symmetry. Various features of this kind of solutions of
Einstein-Maxwell equations have been considered
\cite{Deh3,Dehkhod}. Here, we want to investigate the influence of
rotation on the vortex solution of Einstein-Maxwell-Higgs
equations. Since an analytic solution to the Abelian Higgs field
equations appears to be intractable, we confirm by numerical
calculations that the rotating charged black string can be dressed
by Abelian Higgs field as hair.

In Sec. \ref{Vor}, we consider the Abelian Higgs field equations
in the background of a rotating charged black string. Section
\ref{Num} is devoted to the numerical solutions of the field
equations for different value of the rotation parameter and
winding number. In Sec. \ref{Self}, by studying the behavior of
the Abelian Higgs field energy-momentum tensor, we find the effect
of the vortex self-gravity on the rotating charged black string
background. We append some closing remarks in Sec. \ref{clos}.

\section{Field Equations\label{Vor}}

The Lagrangian of Einstein gravity in the presence of
electromagnetic and Abelian Higgs fields is \cite{Cha1}
\begin{equation}
\mathcal{L}=\mathcal{R}\text{ }-2\Lambda -F_{\mu \nu }F^{\mu \nu }+\mathcal{L%
}_{H},  \label{Lag}
\end{equation}
where $\Lambda =-3/l^{2}$ is the negative cosmological constant
and $F_{\mu \nu }=\partial _{\mu }A_{\nu }-\partial _{\nu }A_{\mu
}$ is the field strength tensor associated with the
electromagnetic field $A_{\mu }$. The first three terms in Eq.
(\ref{Lag}) are the Einstein-Maxwell volume terms with negative
cosmological constant, and $\mathcal{L}_{H}$ is the Lagrangian for
Abelian Higgs system minimally coupled to gravity given as
\begin{equation}
\mathcal{L}_{H}=-\frac{1}{2}(\mathcal{D}_{\mu }\Phi )^{\dagger }\mathcal{D}%
^{\mu }\Phi -\frac{1}{16\pi }\mathcal{F}_{\mu \nu }\mathcal{F}^{\mu \nu
}-\xi (\Phi ^{\dagger }\Phi -\eta ^{2})^{2}.  \label{LagH}
\end{equation}
The matter content of the Abelian Higgs system consists of the
complex Higgs
field, $\Phi $, and a $U(1)$ gauge field $B_{\mu }$ with strength $\mathcal{F%
}_{\mu \nu }=\partial _{\mu }B_{\nu }-\partial _{\nu }B_{\mu }$.
Both the Higgs scalar and the gauge field become massive in the
broken
symmetry phase. They are coupled through the gauge covariant derivatives $%
\mathcal{D}_{\mu }=\nabla _{\mu }+ieB_{\mu }$, where $\nabla _{\mu }$ is the
spacetime covariant derivative. We use units in which $8\pi G=c=1$.

The fields in $\mathcal{L}_{H}$ will be treated as ``test field'',
i.e., their energy momentum tensor is supposed to yield a
negligible contribution to the source of the gravitational field.
Notice that we have two different gauge fields, $F$ and
$\mathcal{F}$, and each is treated in a different manner. It is
only $\mathcal{F}$ that couples to the Higgs scalar field and is
therefore subject to spontaneous symmetry breaking. The other
gauge field, $F$, can be thought of as the free massless Maxwell
field which apart from modifying the background geometry, its
dynamic will be of little concern to us here. It is conventional
to express the field equations in terms of real fields $X(x^{\mu
})$, $\omega (x^{\mu })$, and $P_{\mu }(x^{\nu })$ defined as
\begin{eqnarray}
\Phi (x^{\mu }) &=&\eta X(x^{\mu })e^{i\omega (x^{\mu })},  \label{Phfield}
\\
B_{\mu }(x^{\nu }) &=&\frac{1}{e}[P_{\mu }(x^{\nu })-\nabla _{\mu }\omega
(x^{\mu })],  \label{XPfield}
\end{eqnarray}
where $X$ and $P_{\mu }$ are the scalar Higgs field and massive vector
boson, and $\omega $ is a gauge degree of freedom. In terms of these new
variables, the Lagrangian, $\mathcal{L}_{H}$ and the field equations can be
written as
\begin{eqnarray}
&& \mathcal{L}_{H}(X,P_{\mu }) =-\frac{\eta ^{2}}{2}(\nabla _{\mu
}X\nabla ^{\mu }X+X^{2}P_{\mu }P^{\mu })-\frac{1}{16\pi
e^{2}}F_{\mu \nu }^{\prime
}F^{\prime \mu \nu }-\xi \eta ^{4}(X^{2}-1),  \label{Fi1} \\
&&\nabla _{\mu }\nabla ^{\mu }X-XP_{\mu }P^{\mu }-4\xi \eta ^{2}X(X^{2}-1)=0,
\\
&&\nabla _{\mu }F^{\prime \mu \nu }-4\pi e^{2}\eta ^{2}P^{\nu }X^{2}=0,
\label{Fi2} \\
&&G_{\mu \nu }-\frac{3}{l^{2}}g_{\mu \nu }=T_{\mu \nu }^{em}+\mathcal{T}%
_{\mu \nu }^{Higgs},  \label{Fi3}
\end{eqnarray}
where $F^{\prime \mu \nu }=\nabla ^{\mu }P^{\nu }-\nabla ^{\nu
}P^{\mu }$ is the field strength of the corresponding gauge field
$P^{\mu }$, and $T_{\mu \nu }^{em}$ and $\mathcal{T}_{\mu \nu
}^{Higgs}$ are the stress energy tensors of the electromagnetic
and Higgs fields given by
\begin{eqnarray}
&&T_{\mu \nu }^{em}=2F_{\mu }^{\sigma }F_{\sigma \nu }-\frac{1}{2}%
F^{2}g_{\mu \nu },  \label{Elstr} \\
&&\mathcal{T}_{\mu \nu }^{Higgs}=\eta ^{2}\nabla _{\mu }X\nabla _{\nu
}X+\eta ^{2}X^{2}P_{\mu }P_{\nu }+\frac{1}{4\pi e^{2}}\mathcal{F}_{\mu
\sigma }\mathcal{F}_{v}^{\sigma }+g_{\mu \nu }\mathcal{L}_{H}.  \label{Hstr}
\end{eqnarray}

Note that the real field $\omega $ is not itself a physical quantity.
Superficially it appears not to contain any physical information. However,
if $\omega $ is not single valued this is no longer the case, and the
resultant solutions are referred to as vortex solutions \cite{NO}. In the
presence of vortex, the requirement that the $\Phi $ field be single valued
implies that the line integral of $\omega $ over any closed loop is $\pm
2\pi N$, where $N$ is the winding number of the vortex. In this case the
flux of the field $\mathcal{F}$ passing through such a closed loop is
quantized with quanta $2\pi /e$.

It must be noted that no exact solution are known for the coupled
Einstein-Higgs equations (\ref{Fi1})-(\ref{Fi3}) even in
situations having both no electromagnetic charge and no horizon.
Thus we should try to solve these coupled differential equations
approximately. In the first order approximation, we solve Eqs.
(\ref{Fi1}) and (\ref{Fi2}) in the background of charged rotating
black string and then we will carry out numerical calculations to
solve Eq. (\ref{Fi3}). The metric of the charged rotating black
string can be written as \cite{Lem1, Deh3}
\begin{equation}
ds^{2}=-\Gamma \left( \Xi dt-ad\phi \right) ^{2}+\frac{r^{2}}{l^{4}}\left(
adt-\Xi l^{2}d\phi \right) ^{2}+\frac{dr^{2}}{\Gamma }+\frac{r^{2}}{l^{2}}%
dz^{2},  \label{met2}
\end{equation}
where
\begin{equation}
\Gamma =\frac{r^{2}}{l^{2}}-\frac{bl}{r}+\frac{\lambda ^{2}l^{2}}{r^{2}},%
\hspace{0.5cm}A_{\mu }=-\frac{\lambda l}{r}\left( \Xi \delta _{\mu
}^{0}-a\delta _{\mu }^{\phi }\right) ,\hspace{0.5cm}\Xi =\sqrt{1+a^{2}/l^{2}}%
.  \label{Fg}
\end{equation}
$b$ and $\lambda $ are the constant parameters of the metric which are
related to the mass and charge per unit length of the black string by $M=b/4$
and $Q=\lambda /2$, and $a$ is the rotation parameter. It is remarkable to
note that the metric (\ref{met2}) reduces to AdS spacetime as $t$ goes to
infinity in contrast to the Kerr-AdS metric, for which the surface at
infinity is also rotating. The horizon of this black string has cylindrical
symmetry that has two inner and outer horizon at $r_{-\text{ }}$ and $r_{+}$%
, provided that $b$ is greater than $b_{crit}$ given by
\begin{equation}
b_{crit}=4\times 3^{-3/4}\lambda ^{3/2}.  \label{bcrit}
\end{equation}
In the case that $b=b_{crit}$, we will have an extreme black string.

We seek a cylindrically symmetric solution for the Higgs field equations (%
\ref{Fi1}) and (\ref{Fi2}) in the background of a charged rotating black
string. Thus, we assume that the fields $X$ and $P_{\mu }$ are functions of $%
r$. For the case of vanishing rotation parameter, one can choose
the gauge field as $P_{\mu }\left( r\right) =\left(
0,0,Np(r),0\right) $. The field equations (\ref{Fi1}) and
(\ref{Fi2}) reduce in that case to two equations for the
two unknown functions $X(r)$ and $P(r)$. In the present case the field equations (\ref{Fi1}%
) and (\ref{Fi2}) reduce to three equations and one may therefore
use the gauge choice
\begin{equation}
P_{\mu }\left( r\right) =\left( S(r),0,NP(r),0\right) .  \label{P}
\end{equation}
The field equations (\ref{Fi1}) and (\ref{Fi2}) reduce to
\begin{eqnarray}
&&r^{2}{\Gamma }\,X^{\prime \prime }+\left( 4l^{-2}\,r^{3}-bl\right)
X^{\prime }-4\,r^{2}X\left( X^{2}-1\right) -X\left( aS+\Xi NP\right) ^{2}-
\nonumber \\
&&-{{r}^{2}\Gamma ^{-1}X\left( \Xi S+Nal^{-2}P\right) ^{2}}=0,  \label{Eqm1}
\\
&&N({r}^{3}{l}^{2}{\Gamma }{\Xi }^{2}-{r}^{5}a^{2}l^{-2})P^{\prime
\prime
}+N(2{r}^{4}+b{l}^{3}r{\Xi }^{2}-2\,{\lambda }^{2}{l}^{4}{\Xi }%
^{2})P^{\prime }+({l}^{2}{\Xi }^{2}-{{r}^{2}a^{2}l^{-2}{\Gamma }^{-1}})\times
\nonumber \\
&&N\alpha {r}^{3}X^{2}P+{l}^{2}a\Xi ({{\lambda }^{2}{l}^{2}}-{bl{r}}%
)\left( rS^{\prime \prime }-{S}^{\prime }+{\alpha \,rX^{2}{\Gamma }^{-1}S}%
\right) -a\Xi {{\lambda }^{2}{l}^{4}S}^{\prime }=0,  \label{Eqm2} \\
&&{r}^{3}\left( {\Xi }^{2}{r}^{2}-{\Gamma }\,{a}^{2}\right) S^{\prime \prime
}+[2\,{r}^{4}-{a}^{2}({blr}-2\,{{\lambda }^{2}{l}^{2})]}S^{\prime }+\alpha
\,X^{2}{r}^{3}\Gamma ^{-1}\left( {{\Xi }^{2}{r}^{2}}-{\Gamma a}^{2}\right) S
\nonumber \\
&&-N\Xi \,a[\left( rP^{\prime \prime }-{P}^{\prime }+{\alpha r\,X^{2}{\Gamma
}^{-1}P}\right) ({{\lambda }^{2}{l}^{2}}-{bl{r)}}-{{\lambda }^{2}{l}^{2}P}%
^{\prime }]=0,  \label{Eqm3}
\end{eqnarray}
where $\alpha =4\pi e^{2}/\xi $ and the prime denotes a derivative
with respect to $r$. It is worthwhile to mention that even in the
pure flat or (anti) de Sitter spacetimes no exact analytic
solutions are known for equations (\ref{Fi1}) and (\ref{Fi2}). For
asymptotically AdS spacetimes, one of us have showed that the
Abelian Higgs equations of motion in the background of charged
black string spacetime have vortex solution \cite{Dehjal}. Here we
want to investigate the influence of rotation parameter on the
vortex solutions.

\section{Numerical solutions\label{Num}}

\begin{figure}[tbp]
\epsfxsize=6cm \centerline{\epsffile{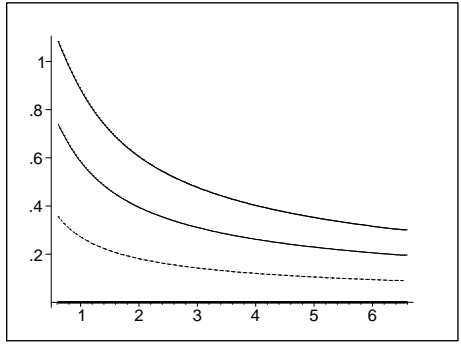}}
\caption{$E_{Higgs}\times 10^{3}$ versus $r$ for $N=1$, $a=0$ (touch the
horizontal axis), $0.25$ (dotted), $0.5$ (solid) and $0.7$ (bold).}
\label{Figure1}
\end{figure}

\begin{figure}[tbp]
\epsfxsize=6cm \centerline{\epsffile{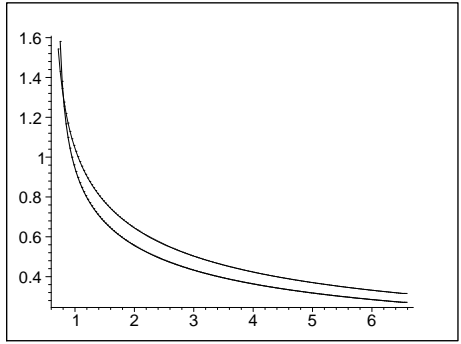}}
\caption{$H_{Higgs}$ versus $r$ for $N=1$, $a=0$ (solid) and $0.7$ (bold).}
\label{Figure2}
\end{figure}

\begin{figure}[tbp]
\epsfxsize=6cm \centerline{\epsffile{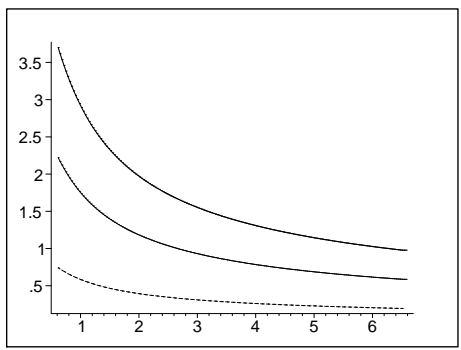}}
\caption{$E_{Higgs}\times 10^{3}$ versus $r$ for $N=1$ (dotted), $3$
(solid), and $5$ (bold).}
\label{Figure3}
\end{figure}

\begin{figure}[tbp]
\epsfxsize=6cm \centerline{\epsffile{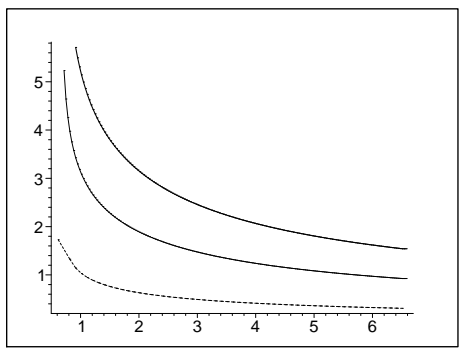}}
\caption{$H_{Higgs}$ versus $r$ for $N=1$ (dotted), $3$ (solid), and $5$
(bold).}
\label{Figure4}
\end{figure}

\begin{figure}[tbp]
\epsfxsize=6cm \centerline{\epsffile{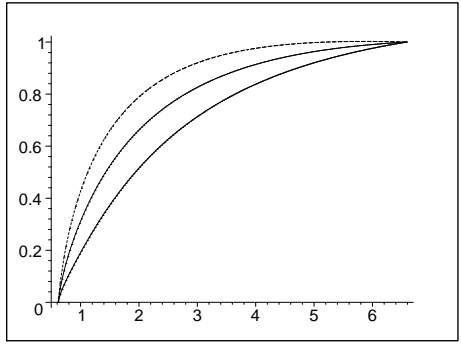}}
\caption{$X(r)$ versus $r$ for $N=1$ (dotted), $3$ (solid), and $5$ (bold).}
\label{Figure5}
\end{figure}

The main difference between this rotating black string and the non rotating
case considered in Ref. \cite{Dehjal} is the following. In the static case
we had only two unknown functions $X(r)$ and $P(r)$, but here we encounter
with three unknown functions $X(r)$, $P(r)$ and $S(r)$. The latter function $%
S(r)$ which is the time component of $B_{\mu }$ creates a radial
electric-type field, $F_{tr}$ that goes to zero as the rotation parameter
goes to zero. It should be noted that even in the case of $b=\lambda =0$, no
exact analytic solutions are known for Eqs. (\ref{Eqm1})-(\ref{Eqm3}). So,
here we seek the existence of vortex solutions for these coupled
differential equations numerically. First, we must take appropriate boundary
conditions. Since at a large distance from the horizon the metric (\ref{met2}%
) reduces to AdS spacetime, we demand that our solutions go to the solutions
of the vortex equations in AdS spacetime given in \cite{Deh1}. This requires
that we demand ($X\rightarrow 1$, $P\rightarrow 0$) as $r$ goes to infinity
and ($X=0$, $P=1$) on the horizon. For consistency with the non-rotating
case \cite{Dehjal}, we take $S=0$ on the horizon. Also one may note that the
electric field, $F_{tr}^{\prime }$ which is proportional to $S^{\prime }$
should be zero as $r$ goes to infinity. We employ a grid of points $r_{i}$,
where $r_{i}$ goes from $r_{H}$ to some large value of $r$ ($r_{\infty }$)
which is much greater than $r_{H}$. We rewrite Eqs. (\ref{Eqm1})-(\ref{Eqm3}%
) in a finite deference language and use the over relaxation method \cite
{Num} to calculate the numerical solutions of $X(r),$ $P(r)$ and $S(r)$ for
different values of the rotation parameter and winding number. The numerical
results of calculations are shown in Figs \ref{Figure1}-\ref{Figure5}.

First, we investigate the influence of rotation parameter on the solutions
of field equations (\ref{Eqm1})-(\ref{Eqm3}). We carry out all the
calculation for $l=1$, $\lambda =0.2$ and $b=0.3$ for which the radius of
horizon is $r_{H}=0.6173$. Figures \ref{Figure1} and \ref{Figure2} show the
behavior of the electric, $E_{Higgs}=F_{tr}^{\prime }$, and the magnetic, $%
H_{Higgs}=F_{\phi r}^{\prime }$, fields associated with the field $P_{\mu }$
respectively for different values of the rotation parameters. We use the
subscript ``Higgs'' for these electromagnetic field to emphasis that they
are coupled with the Higgs scalar field $\Phi $. As one can see in Fig. \ref
{Figure1}, $E_{Higgs}$ is zero for $a=0$, and becomes larger as $a$
increases. The magnetic field $H_{Higgs}$ is plotted in Fig. \ref{Figure2}
for different values of angular momentum. This figure shows that the
variation of $H_{Higgs}$ with respect to the rotation parameter is very
slow. Overall, these figures show that the vortex thickness decreases as the
rotation parameter increases. As we mentioned, in the case of non vanishing
rotation parameter we encounter with the electric field $E_{Higgs}$. One may
compute the source of this electric field through the use of Gauss law
numerically. It is notable that the computation of this electric type charge
for $E_{Higgs}$ shows that it increases as $a$ becomes larger. This is
analogous to the rotating solutions of Einstein-Maxwell equation discussed
in the context of cosmic string theory for which the electric charge of the
string is proportional to the rotation parameter of the string \cite{Lem}.

Next we investigate the influence of the winding number $N$ on the
solutions of field equations (\ref{Eqm1})-(\ref{Eqm3}) for the
case of rotating charged black string. The results for $a=0.5$ are
shown in Figs. \ref {Figure3}, \ref{Figure4} and \ref{Figure5} for
different values of $N$. As in the case of asymptotically flat,
dS, and AdS spacetimes considered in Refs. \cite{Gre} and
\cite{Deh1,Ghez1,Deh2,Dehjal}, increasing the winding number
yields a greater vortex thickness.

The effects of charge or mass parameters on the solutions of field
equations (\ref{Eqm1})-(\ref{Eqm3}) for the case of rotating
charged black string are the same as the nonrotating case, which
is considered in \cite{Deh3}. The extreme black string needs more
investigation. Bonjour et.al. \cite{Bon2} have shown that vortices
of size smaller than a certain fraction of the radius of the
Reissner-Nordstr\"{o}m black hole will definitely pierce the
horizon, whereas vortices thicker than a certain lower bound will
instead wrap the black hole. We will consider the extreme case in
future.

\section{Asymptotic Behavior of the Solutions of Einstein-Maxwell-Higgs
Equation\label{Self}}

\begin{figure}[bp]
\epsfxsize=6cm \centerline{\epsffile{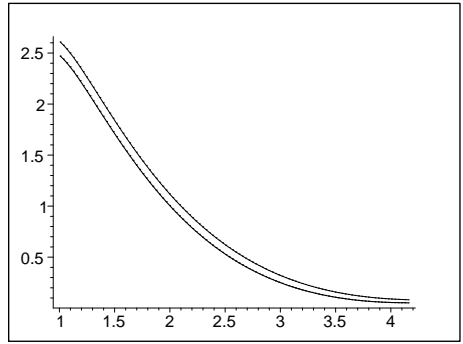}}
\caption{$\mathcal{T}_{t}^{t(0)}$ versus $r $ for $a=0$ (solid)
and $a=0.7$ (bold).} \label{Figure6}
\end{figure}

In previous section we found the solutions of Higgs field equation in the
background of charged rotating black string. Here we want to solve the
coupled Einstein-Maxwell-Higgs differential equation (\ref{Fi1})-(\ref{Fi3}%
). This is a formidable problem even for flat or AdS spacetimes, and no
exact solutions have been found for these spacetimes yet. Indeed, besides
the electromagnetic stress energy tensor, the energy-momentum tensor of the
Higgs field is also a source for Einstein equation (\ref{Fi3}). However,
some physical results can be obtained by making some approximations. First,
we assume that the thickness of the skin covering the black string is much
smaller than all the other relevant length scales. Second, we assume that
the gravitational effects of the Higgs field are weak enough so that the
linearized Einstein-Abelian Higgs differential equations are applicable. We
choose $g_{\mu \nu }\simeq g_{\mu \nu }^{(0)}+\varepsilon g_{\mu \nu }^{(1)}$%
, where $g_{\mu \nu }^{(0)}$ is the rotating charged black string metric in
the absence of the Higgs field and $\ g_{\mu \nu }^{(1)}$ is the first order
correction to the metric. Employing the two assumptions concerning the
thickness of the vortex and its weak gravitational field, the first order
approximation to Einstein equation (\ref{Fi3}) can be written as:
\begin{equation}
G_{\mu \nu }^{(1)}-\frac{3}{l^{2}}g_{\mu \nu }^{(1)}=\mathcal{T}_{\mu \nu
}^{(0)},  \label{Eineq}
\end{equation}
where $G_{\mu \nu }^{(1)}$ is the first order correction to the Einstein
tensor due to $g_{\mu \nu }^{(1)}$ and $\mathcal{T}_{\mu \nu }^{(0)}$ is the
energy-momentum tensor of the Higgs field in the rotating charged black
string background metric with components:
\begin{eqnarray}
\mathcal{T}_{t}^{t(0)}(r) &=&\{-\Gamma l^{4}\,\left( r^{2}\Xi ^{2}-{{\Gamma }%
\,{a}^{2}}\right) S^{\prime ^{2}}-\,\alpha r^{2}l^{4}\,{\Gamma }%
^{2}\,X^{\prime ^{2}}-\,{N}^{2}\Gamma \left( \,\Xi ^{2}l^{4}{\Gamma }-{{a}%
^{2}{r}^{2}}\right) P^{\prime ^{2}}  \nonumber \\
&&-{\alpha \,{N}^{2}X^{2}l}^{4}\Gamma {P^{2}}-\alpha a^{2}\,\left( {\Gamma }%
\,{l}^{2}-{r^{2}}\right) \left( {{N}^{2}P^{2}}-S^{2}l^{2}\right) X^{2}-{%
\alpha r}^{2}l^{4}{\,X^{2}S^{2}}  \nonumber \\
&&-2\alpha r^{2}l^{4}\Gamma \left( X^{2}-1\right) ^{2}\}/2r^{2}l^{4}\Gamma ,
\nonumber \\
\mathcal{T}_{\phi }^{\phi (0)}(r) &=&\{\,\Gamma l^{4}\left( -{{\Gamma }\,{a}%
^{2}}+{r}^{2}{\Xi }^{2}\right) S^{\prime ^{2}}-\,\alpha r^{2}l^{4}\,{\Gamma }%
^{2}\,X^{\prime ^{2}}+\,{N}^{2}\Gamma \left( {\Gamma }\,\Xi ^{2}{l}^{4}-{{a}%
^{2}{r}^{2}}\right) P^{\prime ^{2}}  \nonumber \\
&&+\,\alpha l^{4}\,X^{2}\left( {{N}^{2}{\Gamma }P^{2}}+{{r}^{2}S^{2}}\right)
+\,\alpha \,X^{2}{a}^{2}\left( {\Gamma }\,{l}^{2}-{{r}^{2}}\right) \left( {{N%
}^{2}P^{2}}-{l}^{2}S^{2}\right)  \nonumber \\
&&-2\alpha \Gamma r^{2}l^{4}\,\left( X^{2}-1\right) ^{2}\}/2r^{2}l^{4}\Gamma
,  \nonumber \\
\mathcal{T}_{\phi }^{t(0)}(r) &=&\{2{N}^{2}l^{2}\Gamma a\Xi \,\left( {\Gamma
}\,{l}^{2}-{{r}^{2}}\right) P^{\prime ^{2}}+2\Gamma l^{4}N\left( {{\Gamma }\,%
{a}^{2}}-{r}^{2}{\Xi }^{2}\right) P^{\prime }S^{\prime }-{r^{2}l^{2}PS}
\nonumber \\
&&+2l^{2}NX^{2}\alpha \left( P{a}\left( {\Gamma }\,{l}^{2}-{{r}^{2}}\right)
\left( aS+{\Xi \,NP}\right) \right) \}/2r^{2}l^{4}\Gamma ,  \nonumber \\
\mathcal{T}_{r}^{r(0)}(r) &=&\{\Gamma l^{4}\,\left( {{\Gamma }\,{a}^{2}-\Xi }%
^{2}{r}^{2}\right) S^{\prime ^{2}}+\,\alpha r^{2}l^{4}\,{\Gamma }%
^{2}X^{\prime ^{2}}+\,{N}^{2}\Gamma \left( {\Gamma }\,\Xi ^{2}{l}^{4}-{{a}%
^{2}{r}^{2}}\right) P^{\prime ^{2}}  \nonumber \\
&&+2a\Xi l^{2}\,N\Gamma \left( {{\Gamma l}^{2}}-{r}^{2}\right) P^{\prime
}S^{\prime }-2\alpha r^{2}l^{4}\Gamma \,\left( X^{2}-1\right) ^{2}-\,\alpha
\,X^{2}{a}^{2}\times  \nonumber \\
&&\left( {\Gamma }\,{l}^{2}-{r}^{2}\right) \left( {{N}^{2}P^{2}}+{l}%
^{2}S^{2}\right) -2\alpha \,l^{2}X^{2}Na\Xi \,SP\left( {\Gamma }\,{l}^{2}-{r}%
^{2}\right)  \nonumber \\
&&-\,\alpha l^{4}\,X^{2}\left( {{\Gamma N}^{2}P^{2}}-{{r}^{2}S^{2}}\right)
\}/2r^{2}l^{4}\Gamma ,  \nonumber \\
\mathcal{T}_{z}^{z(0)}(r) &=&\{\Gamma l^{4}\left( -{{\Gamma }\,{a}^{2}}+{r}%
^{2}{\Xi }^{2}\right) S^{\prime ^{2}}-\,\alpha \,r^{2}l^{4}{\Gamma }%
^{2}\,X^{\prime ^{2}}-\,{N}^{2}\Gamma \left( {\Gamma }\,{l}^{4}\Xi ^{2}-{{a}%
^{2}{r}^{2}}\right) P^{\prime ^{2}}  \nonumber \\
&&-2al^{2}\Xi \,N\Gamma \left( {{\Gamma l}^{2}}-{r}^{2}\right) P^{\prime
}S^{\prime }-2\alpha r^{2}l^{4}\Gamma \,\left( X^{2}-1\right) ^{2}-\,\alpha
\,X^{2}{a}^{2}\times  \nonumber \\
&&\left( {\Gamma }\,{l}^{2}-{{r}^{2}}\right) \left( {{N}^{2}P^{2}}+{l}%
^{2}S^{2}\right) -2\alpha l^{2}\,X^{2}Na\Xi \,SP\left( {\Gamma }\,{l}^{2}-{{r%
}^{2}}\right)  \nonumber \\
&&-\alpha \,l^{4}X^{2}\left( {{N}^{2}{\Gamma }P^{2}}-{{r}^{2}S^{2}}\right)
\}/2r^{2}l^{4}\Gamma ,  \label{Thiggs}
\end{eqnarray}
where $X$, $P$, and $S$ are the solutions of the Abelian Higgs
system. As we mentioned in the last section, the vortex thickness
decreases as the rotation parameter increases. This fact is more
clear in Fig. \ref{Figure6} which shows $T_{t}^{t(0)}$ for various
values of $a$.

For convenience, we use the following form, which has cylindrical
symmetry, of the metric
\begin{equation}
ds^{2}=-\widetilde{A}(r)dt^{2}+\widetilde{B}(r)dr^{2}+\widetilde{C}%
(r)dtd\phi +\widetilde{D}(r)d\phi ^{2}+\widetilde{E}(r)dz^{2}.
\label{ABCmetric}
\end{equation}

\begin{figure}[tbp]
\epsfxsize=6cm \centerline{\epsffile{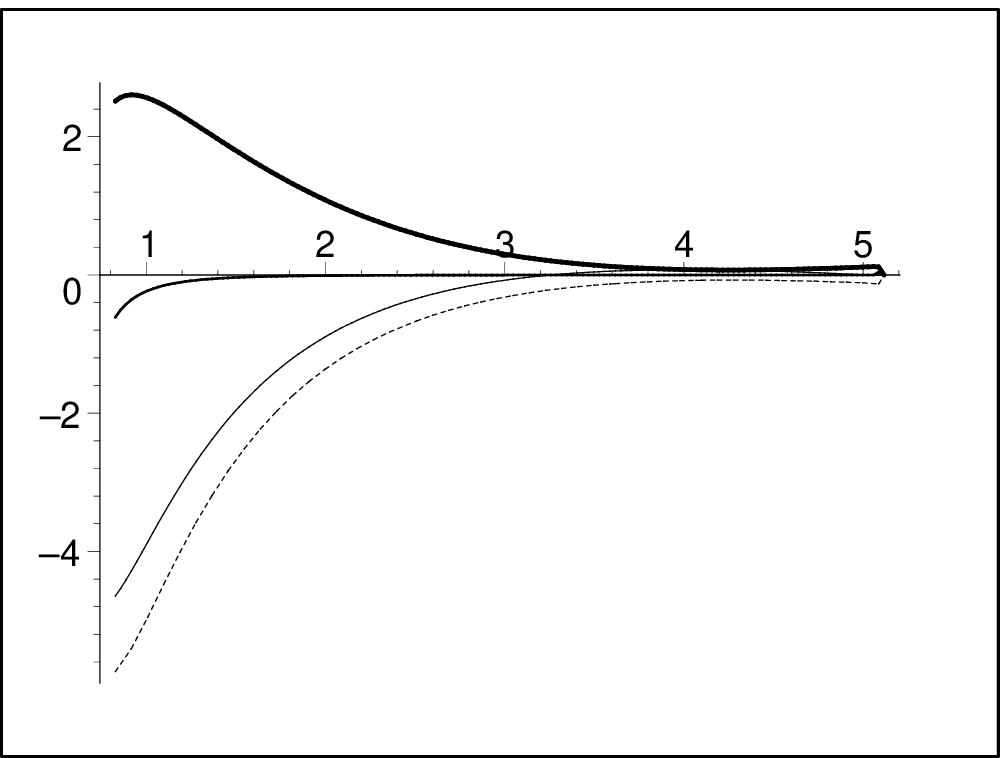}}
\caption{$\mathcal{T}_{t}^{t(0)}=\mathcal{T}_{z}^{z(0)}$ (broken), $\mathcal{%
T}_{\protect\phi }^{\protect\phi (0)}$ (solid), $\mathcal{T}_{\protect%
\phi }^{t(0)}$ (bold) and $\mathcal{T}_{r}^{r(0)}$ (thick-bold), versus $%
r $ for $N=1$, $a=0.5$;.} \label{Figure7}
\end{figure}

\begin{figure}[tbp]
\epsfxsize=6cm \centerline{\epsffile{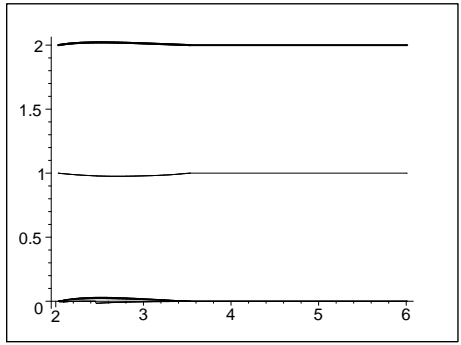}} \caption{$A(r)$,
$B(r)$ and $E(r)$ touch the horizontal axis, $C(r)$ (solid) and
$D(r)$ (bold).} \label{Figure8}
\end{figure}
In order to solve numerically Eq. (\ref{Eineq}), it is better to write the
metric function $\widetilde{A}(r)$ to$\widetilde{\text{ }E}(r)$ as
\begin{eqnarray}
\widetilde{A}(r) &=&A_0(r)[1+\varepsilon A(r)],  \nonumber \\
\widetilde{B}(r) &=&B_0(r)[1+\varepsilon B(r)],  \nonumber \\
\widetilde{C}(r) &=&C_0(r)[1+\varepsilon C(r)],  \nonumber \\
\widetilde{D}(r) &=&D_0(r)[1+\varepsilon D(r)],  \nonumber \\
\widetilde{E}(r) &=&E_0(r)[1+\varepsilon E(r)],  \label{ABCexpa}
\end{eqnarray}
where $A_0(r)=\Gamma \Xi ^2-r^2a^2l^{-4}$, $B_0(r)=\Gamma ^{-1}$, $C_0=2a\Xi
(\Gamma -r^2l^{-2}) $, $D_0(r)=r^2\Xi ^2-\Gamma a^2$, $E_0(r)=r^2l^{-2}$,
yielding the metric of the stationary rotating charged black string in four
dimensions. The Einstein equations (\ref{Eineq}) in terms of the functions $%
A(r)$ to $E(r)$ are given in the appendix. Here we want to obtain
the behavior of these functions for large values of the coordinate
$r$. As one can see from Fig. \ref{Figure7}, the components of the
energy-momentum tensor rapidly go to zero outside the skin, so the
situation is the same as what happened in the static black string
spacetime considered in \cite {Dehjal}. One can solve the
linearized Einstein equation for large values of $r$ numerically.
The results which are displayed in Fig. \ref{Figure8} show that
$A(r)=B(r)=E(r)=0$, and $2C(r)=D(r)=2$. Hence the metric
(\ref{ABCmetric}) can be written as
\begin{equation}
ds^{2}=-A_0(r)dt^{2}+B_0(r)dr^{2}+(1+\varepsilon) C_0(r)dtd\phi
+(1+2\varepsilon) D_0(r)d\phi ^{2}+E_0(r)dz^{2}.  \label{metdef1}
\end{equation}
It is worthwhile to mention that the metric (\ref{metdef1}) is the first
order solution in $\varepsilon$ of the Einstein-Maxwell-Higgs equations far
from the thin string. Of course, one may note that the metric (\ref{metdef1}%
) is the first order approximation of the following metric
\begin{equation}
ds^2=-\Gamma \left( \Xi dt-a\beta d\phi \right) ^2+\frac{r^2}{l^4}\left(
adt-\Xi l^2\beta d\phi \right) ^2+\frac{dr^2}\Gamma +\frac{r^2}{l^2}dz^2,
\label{Metdef}
\end{equation}
which is the exact solution of Einstein-Maxwell gravity. In Eq. (\ref{Metdef}%
), $\beta$ is defined as $\beta=1+\varepsilon$. The above metric describes a
stationary rotating charged black string with a deficit angle $\Delta=2\pi
\varepsilon$. The size of deficit angle $\Delta$ is proportional to $%
2\pi\int r T_t^{t(0)}dr$ \cite{Gre}. Numerical computation shows that the
absolute values of this integral decreases as the rotation parameter
increases, which can also been from Fig. \ref{Figure6}. So, using a physical
Lagrangian based model, we have established that the presence of the Higgs
field induces a deficit angle in the rotating charged black string metric
which decreases as the rotation parameter increases.

\section{Conclusion\label{clos}}

The effect of a vortex on pure AdS spacetime is to create a deficit angle in
the metric in the thin vortex approximation. This fact has been extended to
the charged rotating spacetimes. This is done by including the self-gravity
of the thin vortex in the rotating charged black string background in the
first order approximation discussed in Sec. \ref{Self}. As in the case of
pure AdS \cite{Deh1}, Schwarzschild-AdS \cite{Deh2}, Kerr-AdS, and
Reissner-Nordstr\"{o}m-AdS \cite{Ghez2} spacetimes, we found out that the
effect of a thin vortex on the stationary charged black string is to create
a deficit angle in the metric which decreases as the rotation parameter
increases.

We considered the Abelian Higgs field in the background of a stationary
rotating charged black string. We obtained the numerical solutions for
various values of rotation parameter and found that for a fixed horizon
radius, by increasing the rotation parameter the vortex thickness decrease
very slowly. Also the numerical solutions for various values of winding
number were obtained. These solutions shows that the vortex thickness
increases as winding number increases. These situations hold also in the
case of the extremal black string.

The main difference between the case of Abelian Higgs field in the
background of static black string considered in \cite{Dehjal}, and
this paper is that the time component of the gauge field coupled
to the Higgs scalar field is not zero for non zero rotation
parameter. Indeed, we found that for the case of rotating black
string, there exist an electric field coupled to the Higgs scalar
field. This electric field increases as the rotation parameter
becomes larger. Numerical calculations show that the electric
charge which creates this electric field grows up as the rotation
parameter increases. This is analogous to the results that Dias
and Lemos have found recently for the magnetic rotating string
\cite{Lem}. They showed that the charge per unit length of a
rotating string in the Einstein-Maxwell gravity increases as the
rotation parameter becomes larger, and we found that the electric
charge of the field $F_{\mu \nu }^{\prime }$ coupled to the Higgs
field has the same feature.

Other related problems such as a study of extreme black string in
the presence of Abelian Higgs field, a study of the non-Abelian
vortex solution in asymptotically AdS spacetimes, or other kinds
of fields coupled with gravity remain to be carried out. Work on
these problems is in progress.\newline

\vspace{.5cm}
\noindent {\large \textbf{APPENDIX}}

The Einstein equation mentioned in Sec. (\ref{Self})in terms of the metric
functions $A(r)$ to $E(r)$ are
\begin{eqnarray}
&&{\Xi }^{2}\digamma _{1}(A+D-2C)-4{\lambda }^{2}{l}^{4}a^{2}\Gamma ^{2}({%
\Gamma }\,{\Xi }^{2}{l}^{4}-{r}^{2}a^{2})D+4{\lambda }^{2}{l}^{8}{\Xi }%
^{2}\Gamma ^{2}\left( {\Gamma }\,a^{2}-{r}^{2}{\Xi }^{2}\right) A-{r}%
^{4}l^{2}\Gamma ^{2}\times  \nonumber \\
&&\left\{ 4rl^{4}\,{\Xi }^{2}{\Gamma }+\left( bl^{3}\,-4{r}^{3}a^{2}\right)
\right\} B^{\prime }-{\Xi }^{2}a^{2}{l}^{8}r\Gamma ^{2}\{2\,{b}^{2}r^{2}+{%
\lambda }^{2}(4{\lambda }^{2}{l}^{2}-6\,blr)\}A^{\prime }-12{l}^{4}\,{r}%
^{6}\Gamma ^{2}{B}  \nonumber \\
&&-\,{\Xi ^{2}r}^{2}l^{2}\Gamma \digamma _{{2}}\left( A^{\prime }+D^{\prime
}-2\,C^{\prime }\right) +l^{2}\Gamma ^{2}{r}^{4}\{4rl^{4}{\Gamma }\,{\Xi }%
^{2}+\left( 4r^{3}-bl^{3}\right) \left( a^{2}-l^{2}\right) \}\left(
E^{\prime }+D^{\prime }\right)  \nonumber \\
&&+2\,{\Gamma }^{3}r^{6}l^{6}\,\left( E^{\prime \prime }+D^{\prime \prime
}\right) +\Gamma ^{2}l^{4}a^{2}\{2{\Xi }^{2}{l}^{4}[\,{b}^{2}r+{\lambda }%
^{2}(2{\lambda }^{2}{l}^{2}r^{-1}-\,3bl)]+r^{3}(4{\lambda }^{2}{l}%
^{2}-3blr)\}D^{\prime }  \nonumber \\
&&-2\,{\Xi }^{2}a^{2}{\lambda }^{2}{l}^{6}r\Gamma ^{2}({\Gamma rl}^{2}-r^{3}-%
{bl^{3})}\left( D^{\prime \prime }-C^{\prime \prime }\right)
=4r^{6}l^{6}\Gamma ^{2}\mathcal{T}_{t}^{t(0)},  \nonumber
\end{eqnarray}
\begin{eqnarray}
&&8\Gamma ^{2}(\,{\Gamma l}^{2}-{{r}^{2}}){\lambda }^{2}\Xi \,{l}%
^{6}a^{2}D+8\,{\lambda }^{2}\Xi \,{l}^{6}\Gamma ^{2}a\{l^{2}r^{2}-({\Gamma l}%
^{2}-{{r}^{2}})a^{2}\}C+a{{\Xi }^{3}\digamma _{{3}}\left( A+D-2\,C\right) +}
\nonumber \\
&&r^{4}l^{4}a\Xi \Gamma ^{2}(4rl^{2}{\Gamma }-4{{r}^{3}}+\,{bl}^{3})\left(
B^{\prime }-E^{\prime }-2D^{\prime }\right) -{r}^{2}{l}^{2}\,{a\Xi
\Gamma {\digamma }_{5} \left( A^{\prime }-D^{\prime }\right) }-2{r}^{2}{l}^{2}%
{\Xi a\Gamma \digamma _{{4}}\times }  \nonumber \\
&&{\left( C^{\prime }-D^{\prime }\right) }+2{l}^{4}\Gamma ^{2}\Xi \,a\,\{{l}%
^{5}{r}^{2}\left( {\Xi }^{2}-1\right) {\Gamma }\,\left( {{\lambda }^{2}l}%
-br\right) +{\Xi }^{2}{r}^{6}({r}^{2}-{\Gamma l}^{2})\}\left( D^{\prime
\prime }-C^{\prime \prime }\right) =4r^{6}l^{6}\Gamma ^{2}\mathcal{T}%
_{\varphi }^{t(0)},  \nonumber
\end{eqnarray}
\begin{eqnarray}
&&\digamma {_{{6}}\left( A+D-2\,C\right) }+8\,a^{2}{\lambda }^{2}{l}%
^{6}\Gamma ^{2}\{{\Gamma l}^{2}{\Xi }^{2}-{{r}^{2}\left( -1+{\Xi }%
^{2}\right) \}}\left( C-A\right) +4r^{2}l^{6}{{\lambda }^{2}\Gamma }^{2}{%
\left( \,{\Xi }^{2}l^{2}+a^{2}\right) A+}  \nonumber \\
&&r^{2}\Gamma l^{-4}{{\Xi }^{2}}\,\digamma {_{{7}}\left( A^{\prime
}+D^{\prime }-2\,C^{\prime }\right) }-4{\Xi }^{2}a^{2}{l}^{6}r\Gamma ^{2}\{{b%
}^{2}rl^{2}+{\lambda }^{2}(2\,{\Gamma rl}^{2}-{bl}^{3}-2{r}^{3})\}C^{\prime
}+\,l^{4}r^{2}\Gamma ^{2}{\digamma _{{8}}A^{\prime }}  \nonumber \\
&&+r^{4}l^{4}\Gamma ^{2}({4r}^{3}-bl^{3})E^{\prime }+r^{4}l^{4}\Gamma
^{2}\{4r\,a^{2}{\Gamma }-{\Xi }^{2}({4r}^{3}-bl^{3})\}\left( B^{\prime
}-E^{\prime }\right) +2r^{6}l^{6}\,{\Gamma }^{3}\,\left( E^{\prime \prime
}+A^{\prime \prime }\right)  \nonumber \\
&&-2\,{\Xi }^{2}a^{2}{\lambda }^{2}{l}^{6}r\Gamma ^{2}({\Gamma l}^{2}r-{{r}%
^{3}}-{bl}^{3})\left( A^{\prime \prime }-C^{\prime \prime }\right) -12{l}%
^{4}\,{r}^{6}\Gamma ^{2}{B}=4r^{6}l^{6}\Gamma ^{2}\mathcal{T}_{\varphi
}^{\varphi (0)},  \nonumber
\end{eqnarray}
\begin{eqnarray}
&&{{\Xi }^{2}\digamma _{{9}}\left( A+D-2\,C\right) }-8\,r^{2}{{\lambda }^{2}{%
\Xi }^{2}{l}^{6}\Gamma C}+4r^{2}{\lambda }^{2}{l}^{4}\Gamma {\left( \,{\Xi }%
^{2}l^{2}-a^{2}\right) D}+r^{4}l^{2}\Gamma \left( {4r}^{3}-bl^{3}\right)
E^{\prime }+r^{4}l^{2}\Gamma  \nonumber \\
&&\times \{4r\,a^{2}{\Gamma }-{\Xi }^{2}({4r}^{3}-bl^{3})\}D^{\prime
}+r^{4}\Gamma \{4rl^{4}\,{\Xi }^{2}{\Gamma }-a^{2}({4r}^{3}-bl^{3})\}A^{%
\prime }+\,{\Xi }^{2}a^{2}b{l}^{3}r^{2}\Gamma ({\Gamma l}^{2}-{{r}^{2})}
\nonumber \\
&&\times \left( A^{\prime }+D^{\prime }-2C^{\prime }\right) -{12\,{r}^{6}{l}%
^{2}\Gamma {B}=4r^{6}l^{4}\Gamma }\mathcal{T}_{r}^{r(0)},  \nonumber
\end{eqnarray}
\begin{eqnarray}
&&{{\Xi }^{2}\digamma _{{10}}\left( A+D-2\,C\right) }-16r^{2}{\lambda }^{2}{l%
}^{6}\Gamma ^{2}{\left( \,{\Xi }^{2}l^{2}-a^{2}\right) D}+(32\,{{\lambda }%
^{2}{\Xi }^{2}{l}^{8}r}^{2}\Gamma ^{2}){C}-4{l}^{4}{r}^{4}{\Gamma }^{2}({4r}%
^{3}-bl^{3}){B}^{\prime }+  \nonumber \\
&&4{l}^{4}{r}^{2}{\Gamma }^{2}\{8r^{3}l^{2}\,{\Gamma }+(4{{\lambda }^{2}{l}%
^{2}}-\,{3blr)}\left( {a}^{2}-2l^{2}\right) \}D^{\prime }+4{l}^{4}{r}^{3}{%
\Gamma }^{2}\{8\,{r}^{4}+\,\left( 3\,{\Xi }^{2}-2\right) bl^{2}r-4{{\lambda }%
^{2}{l}^{4}{\Xi }^{2}\}}A^{\prime }  \nonumber \\
&&+4{l}^{2}{r}^{2}{\Gamma }\,{{\Xi }^{2}\digamma _{{11}}\left( A^{\prime
}+D^{\prime }-2C^{\prime }\right) }-8\,{\Xi }^{2}a^{2}{\lambda }^{2}{l}%
^{6}r\Gamma ^{2}\left( rl^{2}{\Gamma }-{r}^{3}-bl^{3}\right) \left(
A^{\prime \prime }+D^{\prime \prime }-2C^{\prime \prime }\right)
+8\,r^{6}l^{6}{\Gamma }^{3}  \nonumber \\
&&\,\times \left( A^{\prime \prime }+D^{\prime \prime }\right) -48{l}%
^{4}r^{6}\Gamma ^{2}{B}=16r^{6}{l}^{6}\Gamma ^{2}\mathcal{T}_{z}^{z(0)},
\nonumber
\end{eqnarray}
where $\mathcal{T}_{\mu }^{\nu (0)}$ is the energy momentum tensor of the
Higgs field given in Eqs. (\ref{Thiggs}) and the functions $\digamma _{i}$'s
are:
\begin{eqnarray}
\digamma _{1} &=&{\Xi }^{4}(32\,{\Gamma }^{3}{l}^{6}{r}^{4}-{\Gamma }^{2}{l}%
^{10}{b}^{2}+{r}^{4}{b}^{2}{l}^{6}-32{\Gamma }\,{l}^{2}{r}^{8}-8\,{\Gamma }%
^{3}{l}^{9}rb+8{\Gamma }^{2}{l}^{7}{r}^{3}b+16\,{r}^{10}-16\,{\Gamma }^{4}{l}%
^{8}{r}^{2}-8{r}^{7}b{l}^{3}  \nonumber \\
&&+8\,{\Gamma }\,{l}^{5}{r}^{5}b)+{\Xi }^{2}({\Gamma }^{2}{l}^{10}{b}%
^{2}+12\,{\Gamma }^{4}{l}^{8}{r}^{2}-28\,{\Gamma }^{3}{l}^{6}{r}^{4}-12\,{%
\Gamma }^{2}{l}^{4}{r}^{6}+60\,{\Gamma }\,{l}^{2}{r}^{8}-32\,{r}^{10}+16\,{r}%
^{7}b{l}^{3}  \nonumber \\
&&-12{\Gamma }\,{l}^{5}{r}^{5}b+4\,{\Gamma }^{3}{l}^{9}rb-8\,{\Gamma }^{2}{l}%
^{7}{r}^{3}b-2\,{r}^{4}{b}^{2}{l}^{6})+4\,{\Gamma }^{4}{l}^{8}{r}^{2}+{r}^{4}%
{b}^{2}{l}^{6}-8\,{r}^{7}b{l}^{3}+4\,{\Gamma }^{3}{l}^{9}rb+16\,{r}^{10}
\nonumber \\
&&+4\,{\Gamma }\,{l}^{5}{r}^{5}b+12\,{\Gamma }^{2}{l}^{4}{r}^{6}-28\,{\Gamma
}\,{l}^{2}{r}^{8}-4\,{\Gamma }^{3}{l}^{6}{r}^{4},  \nonumber \\
\digamma _{2} &=&\,{\Xi }^{4}(4{r}^{7}-4\,{\Gamma }^{3}{l}^{6}r-b{l}^{3}{r}%
^{4}-b{l}^{7}{\Gamma }^{2}+2\,b{l}^{5}{\Gamma }\,{r}^{2}+12{\Gamma }^{2}{l}%
^{4}{r}^{3}-12\,{\Gamma }\,{l}^{2}{r}^{5})+{\Xi }^{2}(2\,{r}^{4}b{l}^{3}-8\,{%
r}^{7}-b{l}^{7}{\Gamma }^{2}  \nonumber \\
&&+16{r}^{5}{\Gamma }\,{l}^{2}-8\,{r}^{3}{\Gamma }^{2}{l}^{4}-{r}^{2}b{l}^{5}%
{\Gamma )}-{r}^{4}b{l}^{3}+4\,{r}^{7}+2\,{\Gamma }^{2}{l}^{7}b-{\Gamma }\,{l}%
^{5}{r}^{2}b+4\,{\Gamma }^{3}{l}^{6}r-4\,{\Gamma }^{2}{l}^{4}{r}^{3}-4\,{%
\Gamma }\,{l}^{2}{r}^{5},  \nonumber \\
&&\digamma _{3}=\,{\Xi }^{2}(32{\Gamma }\,{l}^{2}{r}^{8}-32{\Gamma }^{3}{l}%
^{6}{r}^{4}+8\,{r}^{7}b{l}^{3}-{r}^{4}{b}^{2}{l}^{6}-8\,{\Gamma }^{2}{l}^{7}{%
r}^{3}b-8{\Gamma }\,{l}^{5}{r}^{5}b+8\,{\Gamma }^{3}{l}^{9}rb+{\Gamma }^{2}{l%
}^{10}{b}^{2}-  \nonumber \\
&&16\,{r}^{10}+16\,{\Gamma }^{4}{l}^{8}{r}^{2})+8\,{\Gamma }\,{l}^{5}{r}%
^{5}b+16\,{r}^{10}+32\,{\Gamma }^{3}{l}^{6}{r}^{4}+8\,{\Gamma }^{2}{l}^{7}{r}%
^{3}b-8\,{\Gamma }^{3}{l}^{9}rb-16\,{\Gamma }^{4}{l}^{8}{r}^{2}-32\,{\Gamma }%
\,{l}^{2}{r}^{8}  \nonumber \\
&&+{r}^{4}{b}^{2}{l}^{6}-8\,{r}^{7}b{l}^{3}-{b}^{2}{l}^{10}{\Gamma }^{2},
\nonumber \\
\digamma _{4} &=&\,{\Xi }^{4}(4{r}^{7}-b{l}^{7}{\Gamma }^{2}-b{l}^{3}{r}%
^{4}+12\,{\Gamma }^{2}{l}^{4}{r}^{3}-12\,{\Gamma }\,{l}^{2}{r}^{5}-4\,{%
\Gamma }^{3}{l}^{6}r+2\,b{l}^{5}{\Gamma }\,{r}^{2})+{\Xi }^{2}(8\,{r}^{5}{%
\Gamma }\,{l}^{2}-b{l}^{7}{\Gamma }^{2}  \nonumber \\
&&-4\,{r}^{7}-4{r}^{3}{\Gamma }^{2}{l}^{4}+{r}^{4}b{l}^{3})+2\,{\Gamma }^{2}{%
l}^{7}b+4\,{\Gamma }^{3}{l}^{6}r-4\,{\Gamma }^{2}{l}^{4}{r}^{3},  \nonumber
\\
\digamma _{5} &=&\,{\Xi }^{4}(4{\Gamma }^{3}{l}^{6}r-4{r}^{7}+b{l}^{7}{%
\Gamma }^{2}-2\,b{l}^{5}{\Gamma }\,{r}^{2}-12\,{\Gamma }^{2}{l}^{4}{r}%
^{3}+12\,{\Gamma }\,{l}^{2}{r}^{5}+b{l}^{3}{r}^{4})+\,{\Xi }^{2}(2{r}^{2}b{l}%
^{5}{\Gamma }-b{l}^{7}{\Gamma }^{2}  \nonumber \\
&&+4{r}^{7}+12\,{r}^{3}{\Gamma }^{2}{l}^{4}-12\,{r}^{5}{\Gamma }\,{l}^{2}-4\,%
{\Gamma }^{3}{l}^{6}r-{r}^{4}b{l}^{3})-{\Gamma }\,{l}^{5}{r}^{2}b-4\,{\Gamma
}^{2}{l}^{4}{r}^{3}+4\,{\Gamma }\,{l}^{2}{r}^{5},  \nonumber \\
\digamma _{6} &=&{\Xi }^{6}({b}^{2}{l}^{10}{\Gamma }^{2}+16{\Gamma }^{4}{l}%
^{8}{r}^{2}-32{r}^{4}{\Gamma }^{3}{l}^{6}+32\,{r}^{8}{\Gamma }\,{l}^{2}+8{r}%
^{7}b{l}^{3}+8\,{\Gamma }^{3}{l}^{9}rb-{b}^{2}{l}^{6}{r}^{4}-8\,{r}^{5}{%
\Gamma }\,{l}^{5}b-16\,{r}^{10}  \nonumber \\
&&-8{r}^{3}{\Gamma }^{2}{l}^{7}b)+{\Xi }^{4}{r}^{4}{b}^{2}{l}^{6}+2\,{\Xi }%
^{4}(4b{l}^{3}{r}^{7}-18\,{r}^{8}{\Gamma }\,{l}^{2}+6b{l}^{7}{\Gamma }^{2}{r}%
^{3}-8\,b{l}^{9}{\Gamma }^{3}r+30{r}^{4}{\Gamma }^{3}{l}^{6}+6\,b{l}^{5}{%
\Gamma }\,{r}^{5}  \nonumber \\
&&-4{r}^{6}{\Gamma }^{2}{l}^{4}+8{r}^{10}-{b}^{2}{l}^{10}{\Gamma }^{2}+16{%
\Gamma }^{4}{l}^{8}{r}^{2})+4\,{\Xi }^{2}({r}^{6}{\Gamma }^{2}{l}^{4}-6{r}%
^{4}{\Gamma }^{3}{l}^{6}+\,{r}^{8}{\Gamma }\,{l}^{2}-{r}^{5}{\Gamma }{l}%
^{5}b+2{\Gamma }^{3}{l}^{9}rb  \nonumber \\
&&+4{\Gamma }^{4}{l}^{8}{r}^{2}+{\Gamma }^{2}{l}^{10}{b}^{2}/4)+4(\,{\Gamma }%
^{2}{l}^{4}{r}^{6}-\,{\Gamma }^{3}{l}^{6}{r}^{4}-\,{\Gamma }^{2}{l}^{7}{r}%
^{3}b),  \nonumber \\
\digamma _{7} &=&{\Xi }^{4}(11\,rb{l}^{5}{\lambda }^{4}-10\,{r}^{2}{b}^{2}{l}%
^{4}{\lambda }^{2}-4\,{\lambda }^{6}{l}^{6}+3\,{r}^{3}{b}^{3}{l}^{3})+{\Xi }%
^{2}(3\,{r}^{6}{b}^{2}-7\,{r}^{5}bl{\lambda }^{2}-6\,{r}^{3}{b}^{3}{l}%
^{3}-22\,rb{l}^{5}{\lambda }^{4}  \nonumber \\
&&+8\,{\lambda }^{6}{l}^{6}+4\,{r}^{4}{\lambda }^{4}{l}^{2}+20\,{r}^{2}{b}%
^{2}{l}^{4}{\lambda }^{2})-3\,{r}^{6}{b}^{2}+7\,{r}^{5}bl{\lambda }^{2}-4\,{r%
}^{4}{\lambda }^{4}{l}^{2}+11\,rb{l}^{5}{\lambda }^{4}+3\,{r}^{3}{b}^{3}{l}%
^{3}-4\,{\lambda }^{6}{l}^{6}  \nonumber \\
&&-10\,{r}^{2}{b}^{2}{l}^{4}{\lambda }^{2},  \nonumber \\
\digamma _{8} &=&4\,{\Gamma }\,{l}^{2}{r}^{3}+2\,{l}^{3}{r}^{2}{\Xi }%
^{2}b+8\,{r}^{3}{\Xi }^{2}{\Gamma }\,{l}^{2}-4\,{l}^{3}{\Xi }^{4}b{r}^{2}+8\,%
{\Xi }^{4}{r}^{5}+4\,{l}^{5}{\Xi }^{4}b{\Gamma }-4\,{l}^{5}{\Xi }^{2}b{%
\Gamma }+8\,{\Xi }^{4}{\Gamma }^{2}{l}^{4}r-  \nonumber \\
&&16\,{\Xi }^{4}{\Gamma }\,{l}^{2}{r}^{3}-8\,{\Xi }^{2}{\Gamma }^{2}{l}%
^{4}r+4\,{r}^{5}-{r}^{2}b{l}^{3},  \nonumber \\
\digamma _{9} &=&{\Xi }^{2}(12\,{r}^{4}{\Gamma }^{2}{l}^{4}-24\,{r}^{6}{%
\Gamma }\,{l}^{2}-4\,{l}^{7}b{\Gamma }^{2}r-{l}^{8}{b}^{2}{\Gamma }+4\,{r}%
^{3}b{l}^{5}{\Gamma }+12\,{r}^{8})-16\,{r}^{4}{\Gamma }^{2}{l}^{4}+28\,{r}%
^{6}{\Gamma }\,{l}^{2}+{b}^{2}{l}^{8}{\Gamma }  \nonumber \\
&&-8\,{r}^{3}{\Gamma }\,{l}^{5}b+4\,b{l}^{7}{\Gamma }^{2}r-12\,{r}^{8},
\nonumber \\
\digamma _{10} &=&24\,{\Gamma }^{3}{l}^{9}rb-16\,{\Gamma }^{2}{l}^{7}{r}%
^{3}b+{\Xi }^{2}(76\,{\Gamma }^{3}{l}^{6}{r}^{4}-40\,{\Gamma }^{4}{l}^{8}{r}%
^{2}+20\,{\Gamma }^{2}{l}^{7}{r}^{3}b-48\,{\Gamma }^{2}{l}^{4}{r}^{6}+28\,{%
\Gamma }\,{l}^{2}{r}^{8}-16\,{r}^{10}  \nonumber \\
&&-24\,{\Gamma }^{3}{l}^{9}rb-2\,{\Gamma }^{2}{l}^{10}{b}^{2}-4\,{\Gamma }\,{%
l}^{5}{r}^{5}b-{r}^{4}{b}^{2}{l}^{6}+8\,{r}^{7}b{l}^{3})+40\,{\Gamma }^{4}{l}%
^{8}{r}^{2}-72\,{\Gamma }^{3}{l}^{6}{r}^{4}+44\,{\Gamma }^{2}{l}^{4}{r}^{6}+
\nonumber \\
&&16\,{r}^{10}+4\,{\Gamma }\,{l}^{5}{r}^{5}b-28\,{\Gamma }\,{l}^{2}{r}^{8}+{r%
}^{4}{b}^{2}{l}^{6}-8\,{r}^{7}b{l}^{3}+2\,{b}^{2}{l}^{10}{\Gamma }^{2},
\nonumber \\
\digamma _{11} &=&{\Xi }^{2}(4\,{r}^{7}-{r}^{4}b{l}^{3}-28\,{r}^{3}{\Gamma }%
^{2}{l}^{4}-5\,{r}^{2}b{l}^{5}{\Gamma }+6\,b{l}^{7}{\Gamma }^{2}+8\,{r}^{5}{%
\Gamma }\,{l}^{2}+16\,{\Gamma }^{3}{l}^{6}r)+{r}^{4}b{l}^{3}-8\,{\Gamma }\,{l%
}^{2}{r}^{5}+  \nonumber \\
&&5{\Gamma }{l}^{5}{r}^{2}b-4\,{r}^{7}-6\,{\Gamma }^{2}{l}^{7}b+28\,{\Gamma }%
^{2}{l}^{4}{r}^{3}-16\,{\Gamma }^{3}{l}^{6}r.  \nonumber
\end{eqnarray}

\end{document}